\documentclass{aastex63}

\usepackage{rotating}
\usepackage{xcolor}

\def\s{\phantom}

\def\amin{\ifmmode^{\prime}\else$^{\prime}$\fi}
\def\asec{\ifmmode^{\prime\prime}\else$^{\prime\prime}$\fi}

\def\simgt{\lower.5ex\hbox{$\; \buildrel > \over \sim \;$}}
\def\simlt{\lower.5ex\hbox{$\; \buildrel < \over \sim \;$}}

\newcommand\xte{{\it RXTE\/}}

\newcommand\chandra{{\it Chandra}}

\newcommand\xmm{{\it XMM-Newton}}

\newcommand\swift{{\it Swift\/}}
\newcommand\nustar{\hbox{\it NuSTAR\/}}

\newcommand\tra{SWIFT~J174540.7$ -$290015}
\newcommand\trb{SWIFT~J174540.2$-$290037}
\newcommand\axj{AX J1745.6$-$2901}

\graphicspath{{./}{figures/}}

\shorttitle{X-ray binaries in the Galactic Center}
\shortauthors{Mori et al.}

\begin{document}

\title{The X-ray binary population in the Galactic Center revealed through multi-decade observations}

\author[0000-0002-9709-5389]{Kaya Mori} 
\affiliation{Columbia Astrophysics Laboratory, Columbia University, New York, NY 10027, USA}

\author{Charles~J.~Hailey}
\affiliation{Columbia Astrophysics Laboratory, Columbia University, New York, NY 10027, USA}

\author[0000-0002-7187-9628]{Theo~Y.E.~Schutt}
\affiliation{Columbia Astrophysics Laboratory, Columbia University, New York, NY 10027, USA}

\author[0000-0002-6126-7409]{Shifra~Mandel}
\affiliation{Columbia Astrophysics Laboratory, Columbia University, New York, NY 10027, USA}

\author[0000-0002-6395-4528]{Keri~Heuer}
\affiliation{Department of Physics, Drexel  University, 816 Disque Hall, 32 S. 32nd Street 
Philadelphia, PA 19104, USA}

\author[0000-0002-1323-5314]{Jonathan~E.~Grindlay}  
\affiliation{Harvard-Smithsonian Center for Astrophysics, Cambridge, MA 02138, USA}

\author{Jaesub~Hong} 
\affiliation{Harvard-Smithsonian Center for Astrophysics, Cambridge, MA 02138, USA}

\author[0000-0003-0293-3608]{Gabriele~Ponti} 
\affiliation{INAF - Osservatorio Astronomico di Brera, via E. Bianchi 46, I-23807 Merate, Italy}
\affiliation{Max-Planck-Institut für Extraterrestrische Physik, Giessenbachstrasse, D-85748, Garching, Germany}

\author[0000-0001-5506-9855]{John~A.~Tomsick}
\affiliation{Space Sciences Laboratory, University of California, Berkeley, CA 94720, USA}

\email{kaya@astro.columbia.edu}

\begin{abstract}

We present  {an} investigation of the quiescent and transient X-ray binaries (XRBs) of the Galactic Center (GC). We extended our \chandra\ analysis of the  non-thermal X-ray sources, located in the central parsec,  from \citet{Hailey2018}, using an additional 4.6 Msec of ACIS-S data obtained in 2012--2018. The individual \chandra\ spectra of the 12 sources fit to an absorbed power-law  model with a mean photon index $\Gamma \approx 2$ and show no Fe emission lines. 
Long-term variability was detected from nine of them,  {confirming} that a majority are quiescent XRBs. 
Frequent X-ray monitoring of the GC revealed that the 12 non-thermal X-ray sources, as well as four X-ray transients have shown at most a single  outburst over the last two decades. They are distinct from the six known neutron star LMXBs in the GC, 
which have  {all} undergone multiple outbursts with $\simlt$5 year recurrence time on average.
Based on the outburst history data of  {the broader population of} X-ray transients, we conclude that the 16 sources represent a population of $\sim240$--630 tightly-bound BH-LMXBs with $\sim 4-12$ hour orbital periods, consistent with the stellar/binary dynamics modelling in the vicinity of Sgr A*.  The distribution of the 16 BH-LMXB candidates is disk-like ({\color{black} at 87\% CL}) and aligned with the nuclear star cluster.  Our results have implications for XRB formation and the rate of gravitational wave events in other galactic nuclei.

\end{abstract}


\section{Introduction} \label{sec:intro}

The central parsec of our Galaxy hosts a large population of X-ray binaries (XRBs), as revealed by deep X-ray surveys and long-term monitoring over the last two decades. The discovery of a dozen non-thermal X-ray sources, using 1.4 Msec of \chandra\ ACIS-I observations, suggests the existence of hundreds of quiescent XRBs in the central parsec \citep[][H18]{Hailey2018}.  As an overabundance of X-ray transients  in the GC has been suggested by \citet{Muno2005}, daily \swift/XRT monitoring of a 25\amin$\times$25\amin\ region around Sgr A*, dating back to 2006, has detected a dozen X-ray transients within $\sim20$~pc of the GC \citep{Degenaar2015}. These transients include a variety of source types: a transient magnetar (SGR~J1745$-$29) \citep{Mori2013, Kennea2013, Rea2013}, six NS-LMXBs, which have been identified through the detection of 
type I X-ray bursts, and very faint X-ray transients (VFXTs), whose peak X-ray luminosity is below $10^{36}$~erg\,s$^{-1}$ \citep{Degenaar2012}.  Most recently, two X-ray transients were discovered by \swift\ in 2016, and the follow-up \nustar\ observations suggest that they are  outbursting BH-LMXBs \citep{Mori2019}. 
These LMXBs, observed in quiescent or outbursting states, are more concentrated within a few parsecs from Sgr A*, compared to the magnetic cataclysmic variable (CV)  population, which is spread over the central 10 parsec region \citep{Perez2015, Hailey2016, Hong2016, Zhu2018}. 

These observations of XRBs in the GC not only confirm the prediction that a density cusp of compact objects exists near a supermassive BH \citep{Bahcall1976, Bahcall1977, Morris1993, Miralda2000}, but can also test the fundamental theory of how XRBs are formed in the nuclei of galaxies. 
For example, \citet{Szolgyen2018} suggests that the distribution of isolated 
BHs should be disk-like via vector resonance relaxation around the supermassive BH. \citet{Generozov2018} predicts that the XRB distribution should be concentrated in the central parsec as a result of binary formation via tidal captures, a process enhanced by the high stellar density in the proximity of Sgr A*. 
Further theoretical studies of stellar-mass BH and binary evolution in the Galactic Center reached similar conclusions that the distribution of isolated BHs and binaries should be elongated along the nuclear star cluster \citep{Gruzinov2020} and highly concentrated within $r\sim1$~pc  around Sgr A* \citep{Tagawa2020}.  
Most recently, \citet{Baumgardt2018} and \citet{Panamarev2019} performed N-body simulation to study the dynamics and formation of compact objects in the nuclear star cluster (NSC) around Sgr A*. Besides estimating the rate of tidal disruption and gravitational wave events in the GC, the simulations deduced the density profiles of BHs, white dwarfs and other types of stars in the central few parsec region. 
To examine these theoretical predictions quantitatively, it is vital to identify the XRBs as neutron star (NS) or BH-LMXBs. Detecting more XRBs in the GC will refine their spatial distribution and luminosity function. Ultimately, the number density of BHs and NSs in the GC can be used to normalize the rate of gravitational wave events from stellar remnants in other galactic nuclei \citep{McKernan2018, Fragione2019, Tagawa2020}. 

Despite the growing number of Galactic X-ray transients detected by \swift\ and {\it MAXI}, it is still difficult to positively identify BH systems solely based on their X-ray properties, since NS and BH binaries share many common spectral and timing signatures during outbursts and in quiescence.  Nevertheless, a census of the X-ray outburst history can be used for determining the nature of X-ray transients  \citep{Coriat2012}. Besides some outliers and caveats on source detectability by all-sky X-ray monitors \citep{Knevitt2014, Arur2018, Carbone2019},  NS-LMXB transients  {tend to} have short recurrence times of $\sim$5-–10 years between their X-ray outbursts, while a majority of known BH transients in our Galaxy have undergone outbursts only once in the past 50 years \citep{Corral2016}.

Nearly continuous monitoring of X-ray transients has been carried out only in the past two decades in the GC region. Especially in the central $r < 30$~pc region, daily \swift-XRT monitoring over the last 13 years has detected X-ray transients, including VFXTs, down to $L_X \sim  10^{34}$~erg\,s$^{-1}$. Unlike the more infrequent monitoring of several globular clusters, the \swift/XRT GC observation program provides  the most complete information about the X-ray outburst histories of a concentration of X-ray sources  \citep{Carbone2019}. As a result, it is unambiguously established that all six identified NS-LMXBs within $r \simlt50$ pc and all 5 VFXTs within $r < 10$ pc (where \swift-XRT is sensitive enough to detect VFXTs) have short recurrence times ( {$\simlt 5$ years and }$\simlt 10$ years, respectively) on average\footnote{ {We define the average recurrence time as 20 years (the period of near-constant X-ray monitoring in the GC, starting in 2000) divided by the number of outbursts observed over that period.}}, while other X-ray transients have undergone outbursts only once in the last two decades or longer.  
Additionally, a comparison between the numbers of quiescent and transient BH binaries in the GC can be used to determine the recurrence
time of BH-LMXBs robustly. This is one of the fundamental but highly uncertain parameters in the disk instability models  \citep{Coriat2012}. The X-ray observations of known BH transients in the Galactic plane set only a lower limit  of recurrence time to $\sim50$ years,  with some exceptions, mostly of binaries with  {\color{black} an evolved} donor \citep[e.g., GX~339$-$4,  {V404 Cyg, and V4641 Sgr} with more frequent X-ray outbursts; ][]{Corral2016}. 

This is a follow-up paper to H18, and we investigate the properties of both quiescent and transient XRBs in the GC. 
Based on our subsequent ACIS-S analysis, we update  our spectral and variability studies on the dozen non-thermal X-ray sources discovered by H18 as well as another quiescent XRB (qXRB) candidate (\S\ref{sec:chandra}). Our ACIS-S analysis further establishes the results of H18 by fitting individual source spectra more accurately and detecting variability from four more non-thermal X-ray sources (more specifically, we found an additional  non-thermal X-ray source, and with improved statistics from the addition of ACIS-S data, one of the original dozen H18 sources appears to be a magnetic CV after further analysis).  
We compare their X-ray properties with other known BH or NS transients in our Galaxy and argue that the 12 non-thermal sources, two \swift\ transients in 2016 \citep{Mori2019} and other X-ray transients with single outbursts \citep{Davies1976, Muno2005b} are most consistent with BH-LMXBs  (\S\ref{sec:gc_transients}). Based on recent  correlation studies of X-ray outburst luminosity and recurrence rates, we constrain the orbital period range of the 16 BH-LMXB candidates and discuss the nature of VFXTs in the GC. In \S\ref{sec:bh_distribution}, we investigate the spatial distribution of these 16 BH-LMXB candidates, all located in the central few parsec region. In \S\ref{sec:discussion}, we discuss the spatial distribution, number density and properties of the BH-LMXBs in light of recent theoretical models for XRB formation in the vicinity of Sgr A*. While the paper focuses mainly on studying BH-LMXBs in the GC, a large population of quiescent NS-LMXBs may exist in the GC as observed in globular clusters \citep{Heinke2003}. 
A majority of these sources are undetectable by X-ray telescopes since their dominant soft thermal emission \citep[$kT \simlt 0.1$~keV; ][]{Degenaar2012b, Walsh2015}  is heavily obscured by the large hydrogen column density ($N_{\rm H} \sim 10^{23}$~cm$^{-2}$) in the GC, and they do not show X-ray outbursts due to their low accretion rates. Finally, we summarize our results in \S\ref{sec:summary}. Throughout the paper, we assume a distance to the GC  of 8~kpc  \citep{Camarillo2018}.  


\section{{\it Chandra} ACIS analysis of non-thermal X-ray sources in the GC} \label{sec:chandra} 

In this section, we present our follow-up \chandra\ ACIS analysis of the dozen non-thermal X-ray sources  discovered by H18. We extended the analysis by including 38 additional ACIS-S/HETG observations (3 Msec total exposure) from 2012 and 40 ACIS-S observations (1.6 Msec total exposure) from 2013--2018 and combined them with the ACIS-I data (2002--2011) for the dozen non-thermal X-ray sources published in H18. The improved photon statistics, yielding $\sim 300$--600 net counts for most of the non-thermal X-ray sources, allow us to fit individual source spectra to determine their power-law indices rather than using hardness ratios (\S\ref{sec:chandra_spec}). Additionally, the longer time baseline from incorporating ACIS-I, ACIS-S/HETG and ACIS-S data allows us to detect source variability, thus distinguishing between qXRBs and rotation-powered MSPs (\S\ref{sec:variability}). We did not include ACIS-S observations for one of the non-thermal sources (CXO~J174540.38$-$290033.5) because of the background contamination from the nearby transient magnetar SGR~J1745$-$29.

We used ACIS Extract (AE) software for spectral analysis  \citep{Broos2010}. We extracted source photons from a region encompassing 90\% of the local point spread function (typically $\sim1$\asec) around the \chandra\ position of each source. Background extraction was completed by extracting photons from an annular region centered on the source with a background-to-source region area ratio nominally set to 5 and by avoiding nearby point sources. Response matrices and effective area files were also produced for each observation by AE. We also applied the dust scattering model in XSPEC \textcolor{red}{\citep{Arnaud1996}} to take into account the (energy-dependent) fraction of source photons scattered to outside the extraction region in \chandra\ data \citep{Jin2018}. {\color{black} Note that the long-term degradation of the ACIS effective area (\url{https://cxc.cfa.harvard.edu/ciao/why/acisqecontamN0010.html}) is negligible at $E\simgt2$~keV, where the GC sources are detected, due to the ISM absorption and scattering of soft photons. Hence, the ACIS contamination does not affect our stacked spectral analysis and variability study. } More details can be found in H18. 

\begin{figure*}[ht!]
\begin{center} 
\includegraphics[angle=-90,width=5.83cm]{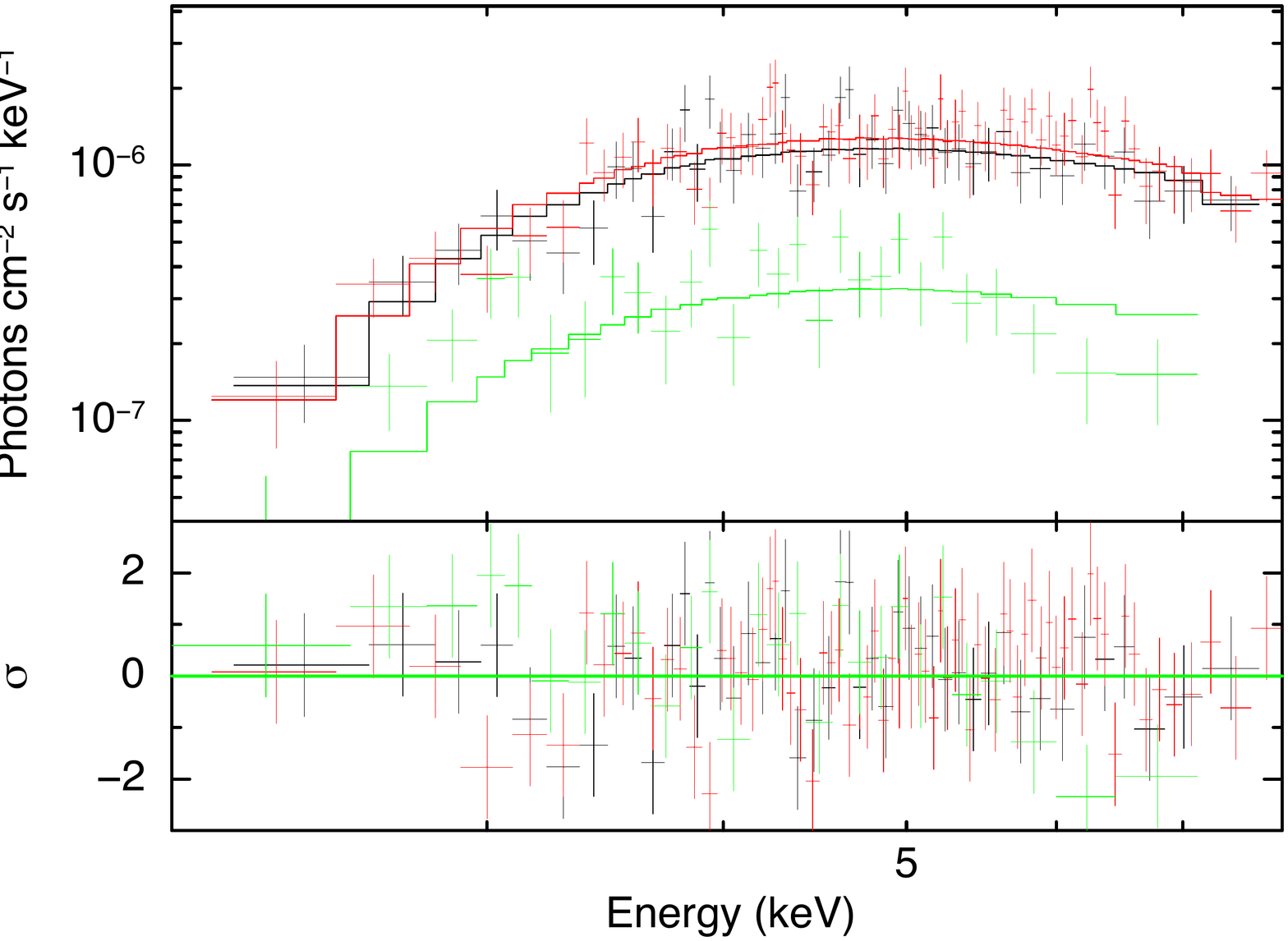}
\includegraphics[angle=-90,width=5.5cm]{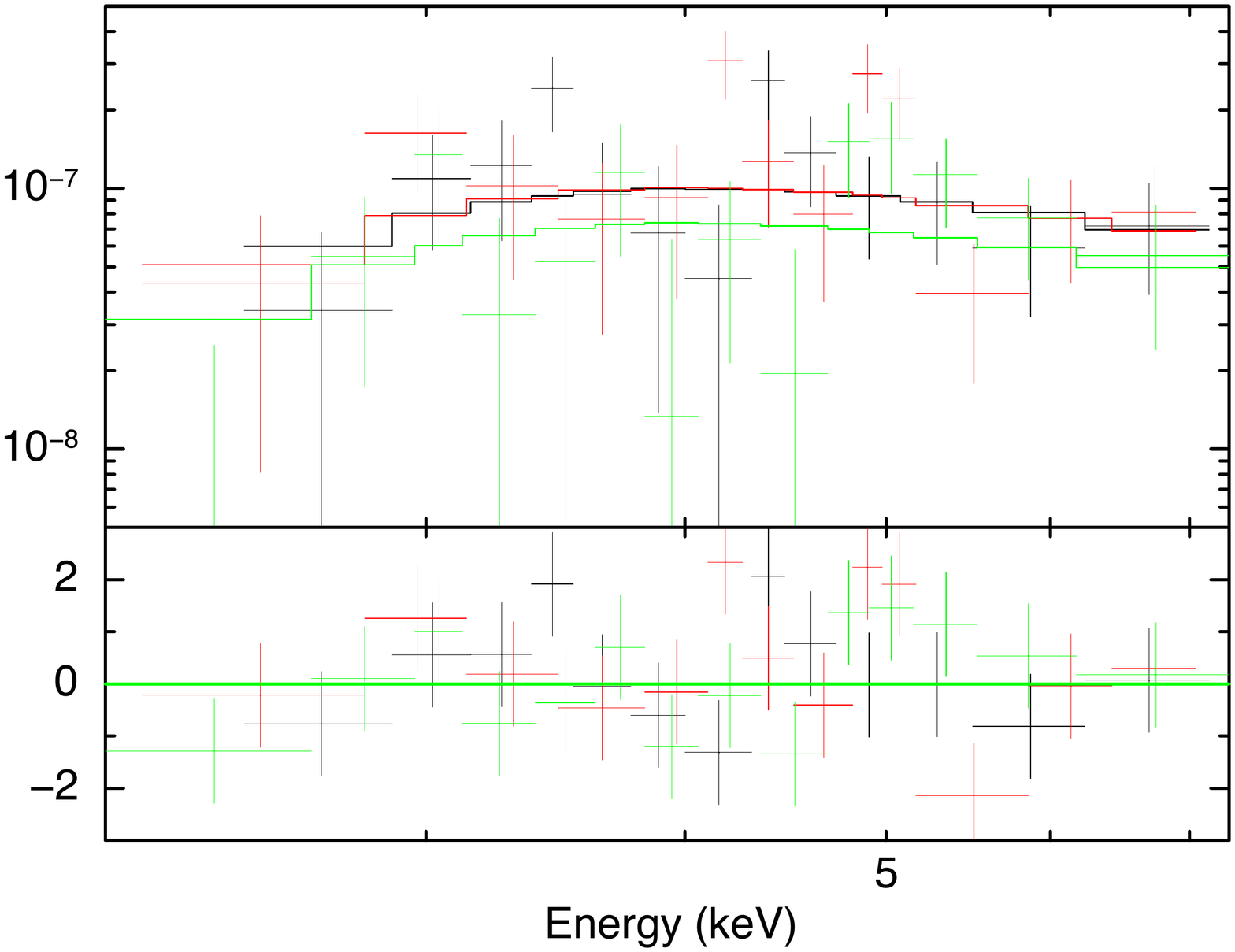}
\includegraphics[angle=-90,width=5.5cm]{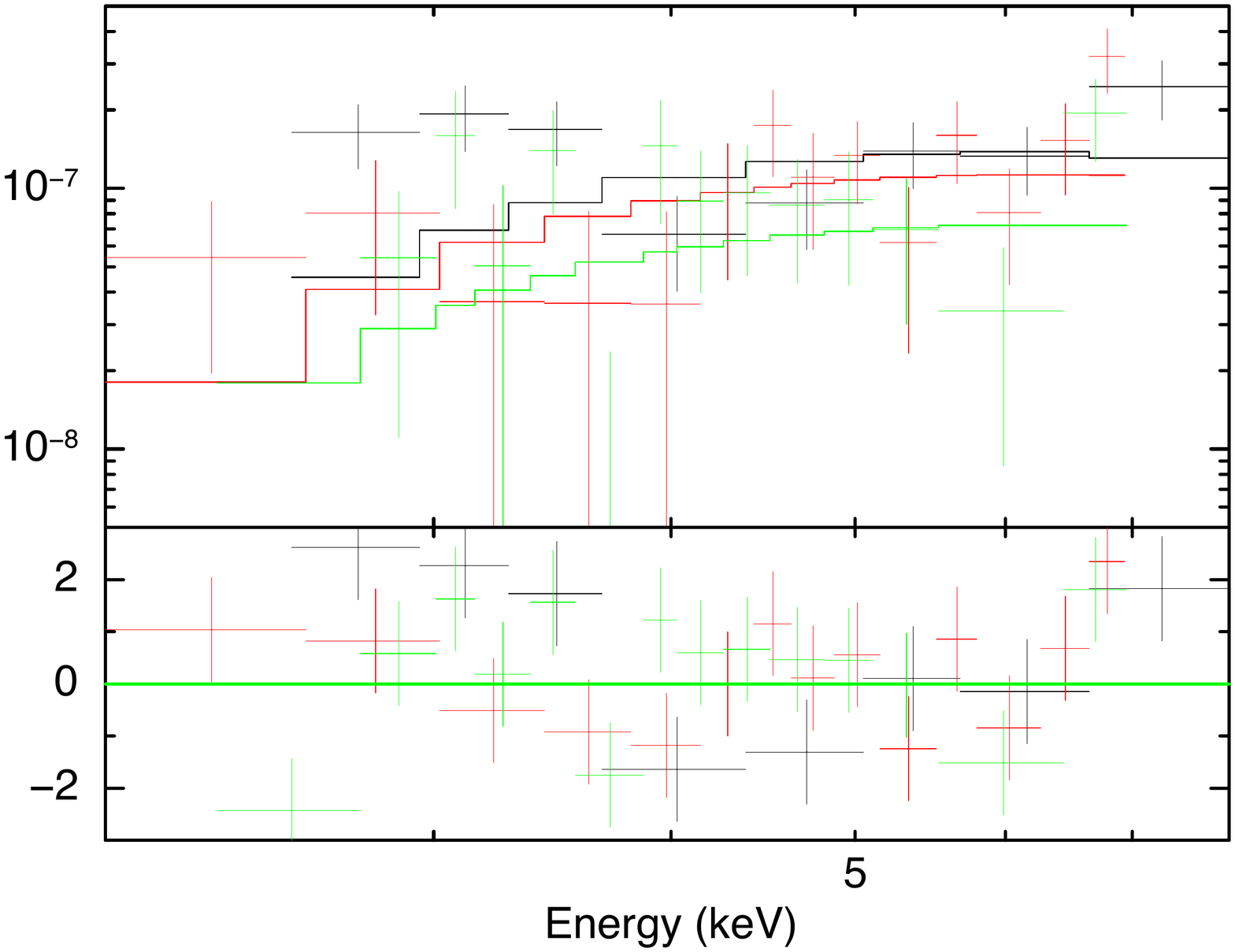}
\caption{\chandra\ ACIS-I (black),  ACIS-S/HETG (red) and ACIS-S (green) spectra of CXO~J174540.16$-$290055.6 (left) and CXO~J174540.37$-$290049.9 (middle), representing the brightest and faintest of the sources, respectively.  {Right: ACIS spectra of CXO~J174539.28$-$290049.1, one of the dozen non-thermal sources listed in H18, {\color{black} which is now classified as a thermal source with atomic lines.}} All ACIS spectra are fitted with an absorbed power-law model; residuals are shown in the lower panels.  
\label{fig:individual_spectra}}
\end{center} 
\end{figure*}

\subsection{Spectral analysis} \label{sec:chandra_spec}

By combining ACIS-I, ACIS-S/HETG and ACIS-S data, we are able to fit individual spectra in XSPEC and determine their power-law indices more accurately. We rebinned each spectrum with 20 counts per bin and applied chi-squared statistics. Fitting unbinned ACIS spectra with C-statistics yielded similar results. 

As a result, we found another qXRB (CXO~J174540.16$-$290055.6) with a non-thermal X-ray spectrum. 
This \chandra\ X-ray source,  {listed in the \citet{Muno2009} and \citet{Zhu2018} catalogs,} is located $\sim$27\asec\ ($\sim$1.1 pc) away from Sgr A*.  In H18, we tagged it as a marginally soft X-ray source {, or qXRB,} potentially with a non-thermal spectrum with $\Gamma \simgt 1.5$, given its hardness ratio HR2$\sim$0.6 (see below for more details). On the other hand, we found that one of the dozen non-thermal sources listed in H18 (CXO~J174539.28$-$290049.1) shows an Fe line feature at $E = 6.7$~keV when we fit its ACIS spectra with an absorbed power-law model,  {yielding a hard photon index $\Gamma = 0.8^{+0.8}_{-0.9}$ ($\chi^2_\nu = 1.9$ for 32 dof). See the right panel of Figure \ref{fig:individual_spectra}. An absorbed thermal APEC model fits the ACIS spectra better with $kT = 7^{+4}_{-3}$~keV with a $\chi^2_\nu = 1.5$ (31 dof). For both models, $N_{\rm H}$ is fixed to $1.2\times10^{23}$~cm$^{-2}$, the best-fit column density from the stacked ACIS spectra of the 12 non-thermal sources (see below). 
{ Based on the presence of Fe line emission and a plasma temperature consistent with the typical range for CVs \citep{Mukai2017}, we conclude that this source is a thermal X-ray source (likely a CV) and exclude it from further analysis.} As we demonstrated with MARX simulations  {in H18}, finding one thermal source out of the dozen non-thermal sources is consistent with the expected rate of false source identification based on the ACIS-I hardness ratio analysis.} All other non-thermal X-ray sources from H18 fit well to an absorbed power-law model with $\chi^2_\nu = 0.8-1.1$ (for $\sim40-150$~dof). We also fit unbinned ACIS spectra using C-statistics and found that the residuals are either insignificant or too narrow (compared to the detector energy resolution), indicating that they are produced by statistical fluctuations.  

{ Based on spectral simulations of mCVs at the GC distance, assuming a two-temperature APEC model with the typical range of plasma temperatures and abundances, we found that mCVs with $>300$ net counts generally have resolvable Fe line emission and significantly harder continuum emission from thermal bremsstrahlung and cyclotron cooling than non-thermal accretion disk emission \citep{Hailey2018}. One exception is GK Per, an IP with a featureless power-law X-ray spectrum with $\Gamma \sim2$, whose Fe abundance was observed to be extremely low \citep[e.g., $A_{\rm Fe} \sim 0.1$ for GK Per;][]{Xu2016}. }
However, we find the existence of a cusp of peculiar GK Per-like CVs in the central parsec region highly unlikely, given that hundreds of thermal X-ray sources in the GC exhibit much higher (and typical for CVs) Fe abundances of $A_{\rm Fe} = 0.7-0.8$ \citep{Muno2004}.  
Thus, we conclude that none of the 12 non-thermal sources can be mCVs based on the additional ACIS-S observations.  In addition, we considered the possibility of non-magnetic CVs (nmCVs), since a fraction of those are bright in the X-ray band \citep{Baskill2005}. We adopted the spectral parameters obtained from 16 typical nmCVs listed in \citet{Xu2016} and simulated \chandra/ACIS spectra using the actual background spectra of the dozen non-thermal sources. \citet{Xu2016} showed that the mean equivalent widths of Fe lines at 6.4, 6.7 and 7.0 keV are 60, 440 and 100 eV, respectively. For completeness, we also considered SS Cyg, which has one of the hardest and brightest X-ray spectra -- with weaker Fe lines -- among nmCVs. Given the number of ACIS counts from each of the sources ($\simgt$ 1,000) and their lack of Fe emission lines, we found that none of the dozen source spectra are consistent with those of typical nmCVs, while one of them could potentially be an SS Cyg-like CV.

Table \ref{tab:13sources} lists the updated spectral properties of the 12 non-thermal X-ray sources. The range in the 2--8~keV luminosity column reflects the flux variation between the ACIS-I, ACIS-S/HETG and ACIS-S spectra. 
For example, we present \chandra/ACIS spectra of CXO~J174540.16$-$290055.6 and CXO~J174540.37$-$290049.9 in Figure~\ref{fig:individual_spectra}, as  examples of bright and faint sources, respectively.  For the new non-thermal source (CXO~J174540.16$-$290055.6), its joint ACIS-I, ACIS-S/HETG and ACIS-S spectra fit well to an absorbed power-law model ($\chi^2_\nu = 0.99$ for 142 dof) with $\Gamma = 2.1\pm0.2$ and $N_{\rm H} = (18^{+2}_{-1}) \times10^{22}$~cm$^{-2}$ (the left panel in Figure~\ref{fig:individual_spectra}). Its 2--8~keV unabsorbed luminosity is $3\times10^{32}$~erg\,s$^{-1}$.  Fitting to an absorbed thermal APEC model, representing typical magnetic CV spectra in the \chandra\ band, led to $N_{\rm H} = (16 \pm 1) \times10^{22}$ cm$^{-2}$, $kT = 7.6 \pm 1$ keV and $A_{\rm Fe} < 0.04$ (90\% C.L.) ($\chi^2_\nu = 0.99$ for 142 dof). The lack of Fe emission lines, as evident from $A_{\rm Fe} \approx 0$, establishes that this is a non-thermal X-ray source, not a CV. 

\begin{figure}[ht!]
\begin{center} 
\includegraphics[angle=-90,width=8.5cm]{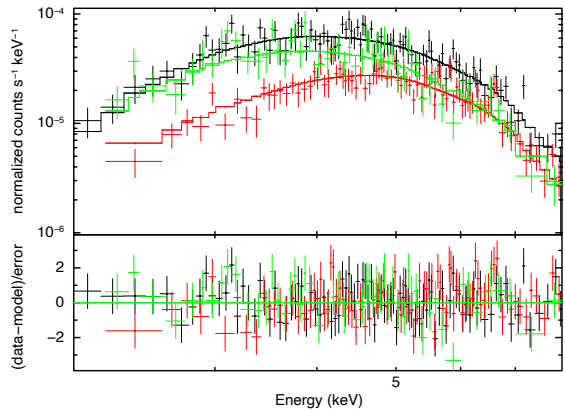}
\caption{Stacked \chandra\ ACIS-I (black), ACIS-S/HETG (red) and ACIS-S (green) spectra of the 12  non-thermal sources jointly fit with an absorbed power-law model. The residuals are shown in the lower panel.   
\label{fig:stacked_spectra}}
\end{center} 
\end{figure}

Subsequently, we stacked ACIS-I, ACIS-S/HETG and ACIS-S  spectra of the 12 non-thermal sources and jointly fit them with an absorbed power-law model by allowing flux normalization factors to vary between the three ACIS spectra (see Figure~\ref{fig:stacked_spectra}).  {The best-fit parameters are $N_{\rm H} = (1.2\pm0.2)\times 10^{23}$~cm$^{-2}$ and $\Gamma = 1.9\pm 0.2$ ($\chi^2_\nu = 1.06$; 234 dof).} 
The lack of Fe lines in the stacked ACIS spectra rules out mCVs whose Fe line EWs exceed $\sim300$~eV \citep{Xu2016}. Moreover,  \citet{Xu2019} found that thermal X-ray sources in the diffuse hard X-ray emission region within  the central 10~pc show strong Fe lines with their total EW above $\sim500$~eV. Clearly, the dozen non-thermal sources exhibit no such strong Fe lines in their individual and stacked ACIS spectra.   

\begin{deluxetable*}{lcccc}[b]
\tablecaption{Spectral and timing properties of the 12 non-thermal X-ray sources}
\tablecolumns{5}
\tablehead{ \colhead{Source name}   & \colhead{Photon index}  & 
\colhead{$L_{\rm X}$ [$10^{31}$ erg\,s$^{-1}$]}  & 
\colhead{Variability significance [$\sigma$]\tablenotemark{a}} &  \colhead{Variability detection methods\tablenotemark{b}} }
\startdata   
174539.87$-$290034.2   & $2.2^{+0.8 }_{-0.7}$ & 8.1--30 & 12.8 & BB (ACIS-I), FV, KS (ACIS-S)     \\
174540.38$-$290033.5   & $1.7^{+1.5}_{-1.8}$\tablenotemark{c} & 0.3--4.9 & 3.8 & FV \\
174540.40$-$290024.1   & $2.0^{+0.8}_{-0.7}$   & 7.9--16 & 4.5 &  FV\\
174540.45$-$290036.3  & $1.5^{+0.7}_{-0.8} $\tablenotemark{c} & 4.0--6.8 & 2.5 & BB (ACIS-I), FV\\
174540.79$-$290024.5   & $2.8^{+1.2}_{-1.0} $ & 4.7--18 & 7.2 & BB (ACIS-I), FV \\
174539.40$-$290040.9   & $2.9^{+1.0}_{-0.9} $   & 8.2--9.8 & 1.6 & BB (ACIS-I)  \\
174540.95$-$290031.2   & $2.1\pm0.7 $\tablenotemark{c}   & 2.0--6.2 & 3.8 & FV  \\
174541.03$-$290026.8   & $1.7^{+0.7}_{-0.8}$\tablenotemark{c} & 3.3--5.2 & 1.6 &  FV, KS (ACIS-I) \\
174540.63$-$290013.4  & $2.1^{+0.9}_{-0.8} $ & 2.3--15 & 12.1 & FV, KS (ACIS-S/HETG)\\
174539.48$-$290045.8  & $2.9^{+0.6}_{-0.5} $   & 6.3--30 & 19.5 & BB (ACIS-I), FV\\
174540.37$-$290049.9   & $2.2\pm0.5 $\tablenotemark{c} & 3.1--4.3 & 1.4 & FV    \\
174540.16$-$290055.6  & $2.0\pm0.4 $    & 20--77 & 24.0 & BB (All), FV, KS (All)
\enddata
\tablecomments{$L_{\rm X}$ for each source indicates a range of unabsorbed 2--8~keV luminosity measured from ACIS-I and ACIS-S spectra.}
\tablenotetext{a}{We listed the highest significance of variability detection among the three methods applied to each source.}
\tablenotetext{b}{We listed the methods which detected  source variability at $>90$\% significance. BB, KS and FV stand for Bayesian Block analysis, KS test and flux variation between ACIS-I, ACIS-S/HETG and ACIS-S spectra, respectively. }
\tablenotetext{c}{The power-law index was not well constrained when $N_{\rm H}$ was fit freely. We fixed $N_{\rm H}$ to $1.2\times10^{23}$~cm$^{-2}$, which is the best-fit column density from the stacked ACIS spectra of the 12 non-thermal sources. }
\label{tab:13sources}
\end{deluxetable*}

\subsection{Source variability} \label{sec:variability}

Source variability is a robust diagnostic to distinguish between LMXBs and rotation-powered millisecond pulsars (rMSPs). Many LMXBs in the quiescent state have shown evidence of X-ray flux variability, by a factor of 2–-5 over a timescale of days to years, due to varying accretion rate \citep{Plotkin2013}. 
While rMSPs also have non-thermal X-ray emission, they display no such long term variability \citep{Bogdanov2006}.

\begin{figure}[ht!]
\begin{center} 
\includegraphics[width=9cm]{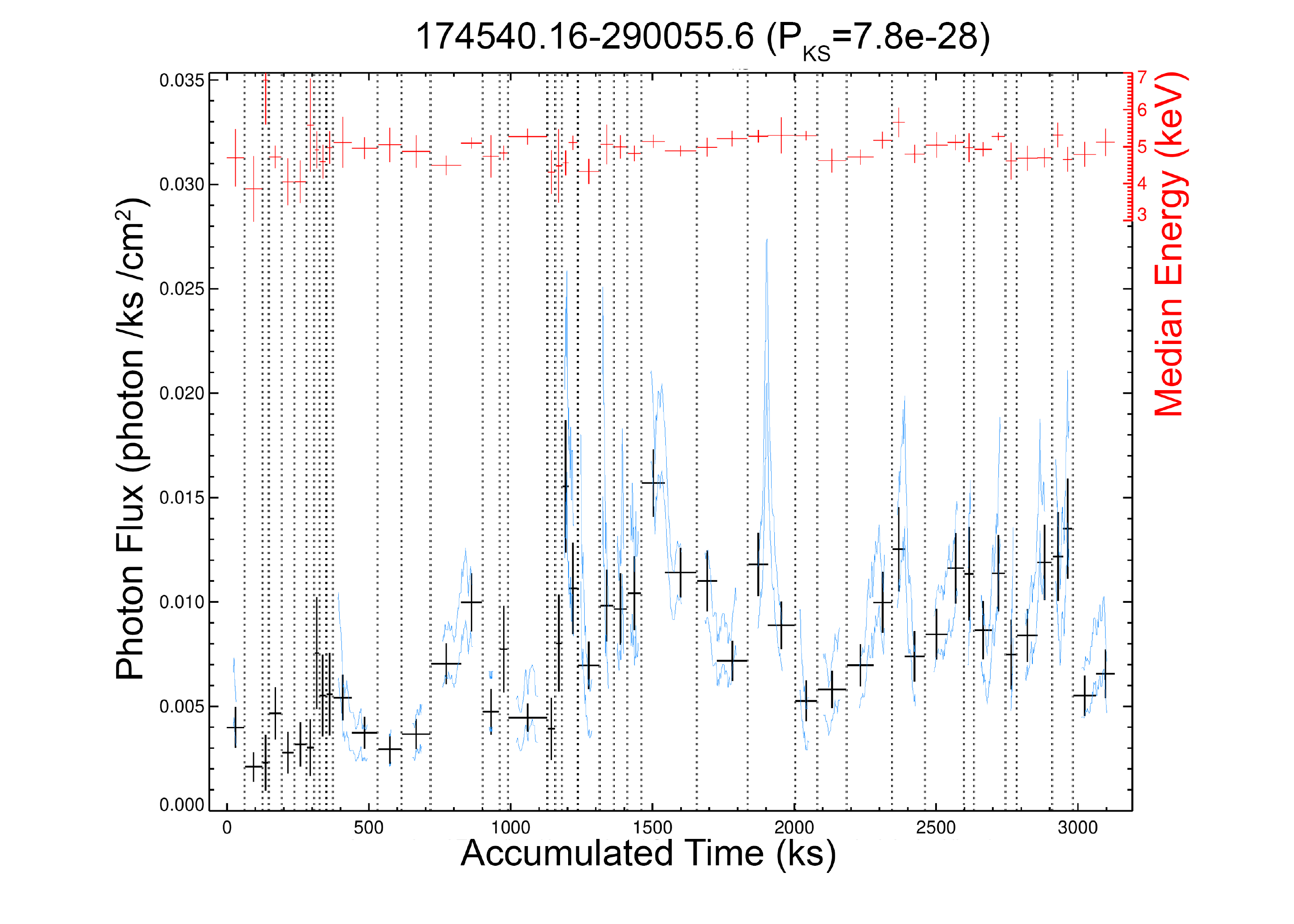}
\caption{\chandra\ ACIS-S sequenced lightcurve of  CXO~J174540.16$-$290055.6, with gross flux in time bins of constant significance (black), median energy of the events in each bin (red), and pointwise 68\% confidence band for the continuous lightcurve (blue). Each observation is separated by a vertical dashed line.  
\label{fig:lc}}
\end{center} 
\end{figure}

We investigated the source variability in three different approaches as listed in Table \ref{tab:13sources} - Bayesian Block (BB) analysis \citep{Scargle2013}, Kolmogorov-Smirnov (KS) test, and flux variation between the ACIS-I (2002--2011), ACIS-S/HETG (2012) and ACIS-S (2013-2018) spectra. 
Following H18, we first combined individual event files in each of the ACIS-I, ACIS-S/HETG and ACIS-S data sets separately, then applied the BB and KS tests to the unbinned photon arrival times in each merged event file. We found that 7 (2) sources were variable at $>$90\% ($>$99\%) confidence level (CL), including the new non-thermal source CXO~J174540.16$-$290055.6. 
Neither BB nor KS tests detected significant variability in the ACIS-S/HETG and ACIS-S data of the other non-thermal sources. 
As presented in \S\ref{sec:chandra_spec}, we jointly fit the ACIS-I, ACIS-S/HETG and ACIS-S spectra for each source with an absorbed power-law model and calculated the flux normalization factor between the three spectra. Nine out of the 12 non-thermal X-ray sources showed X-ray flux variation above 90\% CL. 
For example, in Figure \ref{fig:lc}, \chandra\ 2--8 keV lightcurves of CXO~J174540.16$-$290055.6 show that the source is highly variable with KS test p-value $= 8\times10^{-28}$. The source variability rules out the possibility that this source is a rMSP. We also note that none of the 12 non-thermal sources have had X-ray outbursts ($L_X \simgt 10^{34}$~erg\,s$^{-1}$) detected, despite the frequent GC monitoring over the past two decades.


\section{X-ray binaries in the GC, their outburst history and quiescent states} \label{sec:gc_transients}

Unlike the detection of type I X-ray bursts or pulsations from NS transients, there is no ``smoking gun" for identifying BH transients solely based on their X-ray properties, {\color{black} with a few exceptions. Broad Fe emission lines, often detected from X-ray transients, can result from relativistic effects near a fast-spinning BH. A high spin value above $a_* \sim 0.7$, measured through broad-band X-ray spectral fitting, establishes the BH transient case \citep{Miller2015}, since observed spin values for NS transients are much smaller -- e.g., $a_* = 0.15$ for the NS-LMXB 4U 1728$-$34 \citep{Sleator2016}. Detection of a bright radio jet during the low/hard state has also been considered a diagnostic for BH transients, since the radio jets from outbursting NS-LMXBs are fainter by a factor of $\sim22$ \citep{Gallo2018}. On the other hand, it is difficult to definitively classify  the variable, non-thermal X-ray sources we discussed in \S\ref{sec:chandra} as NS- or BH-LMXBs, since in quiescence, many of their spectral and timing properties are similar. This leaves open the question of whether the non-thermal X-ray sources are NS- or BH-LMXBs. } 

In this section, we review (1) recent studies on the X-ray outburst history of NS/BH LMXBs (\S\ref{sec:outburst_history}) and (2) the X-ray transients discovered in the GC and their outburst history (\S\ref{sec:gc_history}). Based on the most recent, unbiased analysis of X-ray outbursts, we argue that the dozen non-thermal X-ray sources are consistent with BH-LMXBs (\S\ref{sec:gc_srcs_nature}). Their quiescent X-ray spectra, obtained with our \chandra\ ACIS analysis, provide further support for the BH-LMXB scenario (\S\ref{sec:quiescent}). We also discuss the nature of VFXTs, which may belong to a different class of LMXBs, in the GC (\S\ref{sec:vfxt_nature}).  

\subsection{X-ray outburst history of NS and BH LMXBs} \label{sec:outburst_history}

Recently, it has become more evident that X-ray outburst history can be used as a diagnostic tool to infer the nature of LMXBs \citep{Yan2015, Lin2019}.  Except for several outliers (e.g., Cen X-4),\footnote{Two X-ray outbursts were detected from Cen X-4 in 1969 and 1979 \citep{Conner1969, Matsuoka1980}. Since then, no X-ray outburst has been detected from this NS-LMXB.} most NS  transients have a short recurrence time of $\sim$5--10 years  between their X-ray outbursts \citet{Coriat2012}. 
On the other hand, a majority (73\%) of known BH transients in our Galaxy have had only one outburst in the past $\sim$50 years \citep{Corral2016}. {\color{black} Hereafter, for clarity, we refer to X-ray transients with only one detected outburst as single X-ray outburst transients (SXOTs).} 
In the scheme of the disk instability model  (DIM),  \citet{Coriat2012} offered a plausible explanation for the distinct outburst recurrence trends between NS- and BH-LMXBs to the ``transientness'' parameter (i.e. a ratio of the estimated mass transfer rate $\dot{M}$ to a critical accretion rate $\dot{M}_{cr}$), barring some uncertainties on the $\dot{M}$ measurements and ambiguous source types on whether they contain NS or BHs. 

In general, the poor detectability of faint X-ray transients through all-sky monitoring with \swift-BAT and MAXI hampers the accurate measurement of recurrence times. Overall, VFXTs and most X-ray transients at large distances are subject to this selection effect with all-sky X-ray monitoring. Even though several  globular clusters (e.g. NGC {\color{black} 6388}, Terzan 5) have been monitored by \swift/XRT, \citet{Carbone2019} demonstrated  that  their infrequent observing cadence may lead to missing multiple  X-ray  outbursts  and  VFXTs for a range of duty cycles and outburst durations. 
{\color{black} For instance, 47 Tuc, a globular clusters that is among the most extensively observed by \swift/XRT, has had $\simlt200$~ksec total exposure time, whereas the ongoing \swift/XRT GC monitoring program has accumulated several Msec worth of exposures since 2006 \citep{Degenaar2015}.}  
In addition, source identification may be difficult due to the high background X-ray emission and source confusion in  globular clusters (e.g., if a transient is not quickly followed up by \chandra\ during its outburst). \citet{Carbone2019} performed Monte-Carlo simulation for assessing the \swift/XRT detectability of X-ray transients and determined that the \swift/XRT monitoring of the GC offers the only nearly complete survey of X-ray  outbursts down to the luminosity range of VFXTs ($L_X \sim10^{34}$~erg\,s$^{-1}$).\footnote{\citet{Carbone2019} pointed out that an X-ray transient in the GC may remain undetected if its outburst is periodic and somehow coincides with the annual \swift/XRT observation gap that occurs from November to February. However, we consider such a case to be very unlikely, given the typically stochastic nature of X-ray outburst recurrence time and average outburst duration.}   However, this observation bias primarily affects NS transients, since BH-LMXB outbursts tend to be both significantly brighter and longer in duration than their NS counterparts \citep{Yan2015}.  Therefore, NS-LMXB outbursts are much more likely to go undetected, indicating that the outburst frequency of NS transients may be even higher than currently known.  

This makes it highly unlikely that most of the 12 non-thermal GC sources are NS-LMXBs; due to the near-constant X-ray monitoring of the GC region over the past two decades, any outbursts from these sources would likely have been detected (H18).  While persistently quiescent NS-LMXBs do exist -- many such sources have been identified in globular clusters -- those binaries are believed to undergo little to no accretion; their X-ray emission is predominantly thermal and non-varying, as it originates from the NS surface \citep{Bogdanov2016}. Such an origin can be excluded for our 12 non-thermal sources. {\color{black} In globular clusters, only a few confirmed NS-LMXBs (e.g. { CX3 in Terzan 5}) are known to have significant non-thermal components in their X-ray spectra above $E\sim2$ keV \citep{Bahramian2020}, largely due to their low accretion rates (a result of aging binary populations) and the lack of extensive \chandra\ observations (unlike the GC with $\sim7$~Msec total exposure as presented in this paper). Furthermore, the stellar populations and environments of globular clusters are distinct from those in the Galactic disk and GC \citep{Heinke2003}. Therefore, hereafter we adopt the historical data of GC transients collected over the last two decades as the most reliable sample for identifying the dozen non-thermal sources in the central parsec. }

\subsection{X-ray transients in the GC} \label{sec:gc_history}

In 2006 February, \swift\  initiated a daily XRT monitoring program for detecting X-ray flares from Sgr A* and X-ray transients in the central 25\amin$\times$25\amin\ (60~pc $\times$ 60~pc) region. Since then, \swift\ has detected a dozen  X-ray transients (including some recurrent transients discovered earlier) within $\sim50$~pc of the GC \citep{Degenaar2015}, including 
a new transient magnetar \citep{Mori2013}. 
We compiled all the X-ray transients detected in the central $\sim50$~pc region where X-ray telescopes such as \chandra, \xmm\, and \swift\ are sensitive to detecting X-ray outbursts above $\sim10^{34}$~erg\,s$^{-1}$. In total, there have been 20 X-ray transients detected in the GC.

\begin{figure}
\begin{center} 
\includegraphics[width=8.5cm]{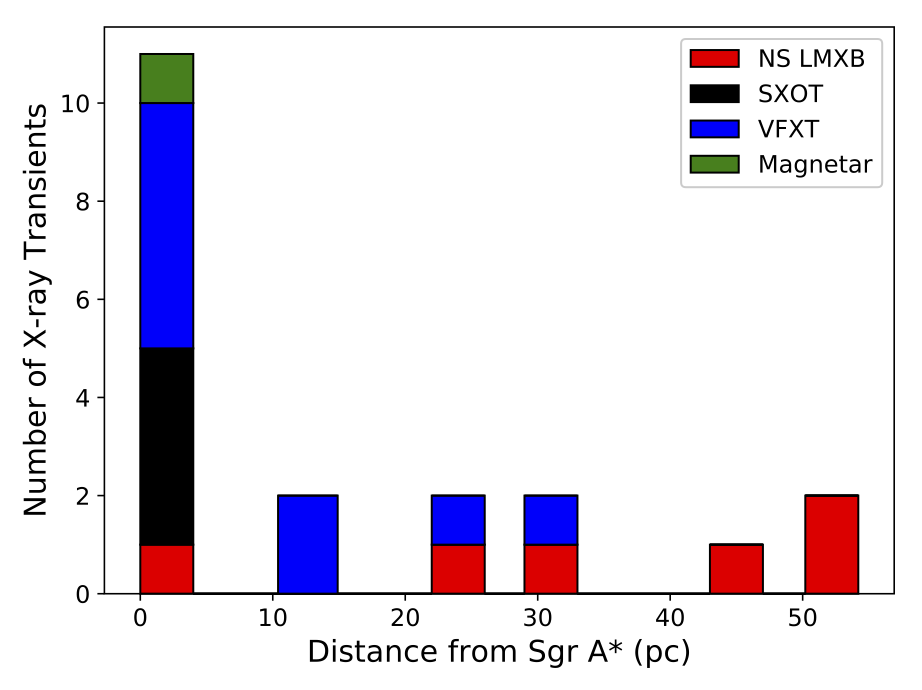}
\caption{Radial distribution of all the X-ray transient sources detected within $r\sim$50~pc of the GC. The number of transients is plotted at different projected distances [pc] from Sgr A*. 5 of the 9 VFXTs are located within $< 10$~pc and recurrent X-ray outbursts have been detected from them by \swift/XRT. 4 other VFXTs, all of which are outside $\sim10$~pc, have been detected only once. 
\label{fig:gc_transients}}
\end{center} 
\end{figure}

Besides the transient magnetar SGR~J1745$-$2900, these GC transients can be classified into three groups: six NS-LMXBs, four SXOTs and 9 VFXTs. The six NS-LMXBs have been identified through type I X-ray bursts and/or pulsation detection \citep{Degenaar2012}. The four SXOTs include the two \swift\ transients detected in 2016 (\tra\ and \trb), CXOGC~J174540.0$-$290031  and 1A~1742$-$289. CXOGC~J174540.0$-$290031 was detected by \chandra\ in 2005 \citep{Muno2005b}. {\color{black} Follow-up \nustar\ observations suggest that the 2016 \swift\ transients are BH-LMXBs due to their high BH spin values of $a_* > 0.9$ \citep{Mori2019}. } An outburst from 1A~1742$-$289, located at $\sim$2~pc from Sgr A*, was detected in the radio and X-ray bands in 1975 \citep{Davies1976}. Its radio position distinguished the source from nearby X-ray transients such as \axj. {\color{black} Bright radio jets in the range of $L_{\rm radio} \sim 10^{32}-10^{33}$~erg\,s$^{-1}$  were observed from both CXOGC~J174540.0$-$290031  and 1A~1742$-$289 during their X-ray outbursts. Following the well-established $L_{\rm radio}$ vs $L_{\rm X}$ correlation in the hard state \citep{Gallo2018}, the radio-loud outbursts suggest that they are BH transients.  Hence, all four SXOTs are likely BH transients, which is consistent with the fact that a majority of known BH transients in the Galactic disk have long recurrence times \citep{Corral2016}.} For the VFXTs in the GC, it is still unknown whether they harbor a NS or BH. 
As shown in Figure~\ref{fig:gc_transients}, about half of all the GC X-ray transients detected by \swift, \chandra\, and \xmm\ are concentrated in the central few parsecs around Sgr A*. All four SXOTs (BH transients) are located at $r \simlt 2$~pc, while the NS-LMXBs and VFXTs are distributed  more widely beyond $r\sim$10~pc.

{\color{black} As mentioned above, the most frequent and sensitive  survey of X-ray transients, down to $L_{\rm X} \sim 10^{34}$~erg\,s$^{-1}$, has been carried out in the GC region \citep{Degenaar2015}.} In the central $r\simlt30$~pc region, daily 
\swift-XRT monitoring of the GC over the last 13 years is believed to have revealed most X-ray transients down to $L_X \sim 10^{34}$~erg\,s$^{-1}$ \citep{Degenaar2015}. Within the central $\sim50$~pc region, all six identified NS-LMXBs have short recurrence times of $\simlt5$ years, and five of the nine  VFXTs have average recurrence times of $\simlt10$ years. 
Note that \swift-XRT may have missed faint/short outbursts from the four non-recurrent VFXTs (like a two-week long $L_X\sim10^{35}$~erg\,s$^{-1}$ outburst detected from SWIFT~J174553.7$-$290347 at $r\sim25$~pc), all of which are located outside 10~pc where the XRT sensitivity decreases rapidly \citep{Degenaar2009}. 

Given the lack of detection of X-ray outbursts in the last two decades, the 12 quiescent, non-thermal X-ray sources, along with the four SXOTs {\color{black} (all of which are likely BH transients)}, are clearly distinct from the 6 known NS-LMXBs and 5 VFXTs within $r < 10$~pc, which have all had recurrent X-ray outbursts. It is also evident that the SXOTs are not VFXTs as  their peak X-ray luminosities\footnote{\citet{Muno2005b} estimated the peak X-ray luminosity of CXOGC  J174540.0$-$290031 exceeded $\sim2\times10^{36}$~erg\,s$^{-1}$ based on the scattered X-ray flux from nearby diffuse X-ray sources.} exceed $L_X \sim 10^{36}$~erg\,s$^{-1}$.

\subsection{Nature of single X-ray outbursters and quiescent XRBs in the GC}
\label{sec:gc_srcs_nature}
 
Besides the GC transient monitoring described above, some recent studies on X-ray outburst history can be used to further infer the nature of the SXOTs and quiescent XRBs in the GC. It is well established that peak X-ray outburst luminosity and recurrence time are related to the physical size of the accretion disk and Roche lobe, and thus to the orbital period, as supported by X-ray observations \citep{Wu2010, Lin2019}. In general, XRBs with shorter orbital periods show fainter X-ray outbursts. \citet{Wu2010} found an empirical relation between the peak $L_X$ and $P_{\rm orb}$ [hours] as $\log{(L_{\rm X}/L_{\rm Edd})} = -1.80+0.64\log{P_{\rm orb}}$, where $L_{\rm Edd}$ is the Eddington luminosity. Note that, for a given $P_{\rm orb}$, BH-LMXBs should (on average) be brighter than NS-LMXBs during X-ray outbursts due to their higher $L_{\rm Edd}$. 

On the other hand, \citet{Lin2019} found  a robust relationship between $P_{\rm orb}$ and recurrence time ($\tau_{\rm rec}$) for bright X-ray outbursts ($L_{\rm X} \simgt  10^{37}$~erg\,s$^{-1}$ at 8 kpc). Their analysis of \swift/BAT, \xte/ASM and {\it MAXI} all-sky monitoring data showed that short-period LMXBs ($P_{\rm orb} \simlt 12$~hours), whether they contain a NS or BH, have undergone only one bright outburst ($L_X \simgt 10^{36}$~erg\,s$^{-1}$) over more than a decade, whereas multiple such outbursts have been observed with $\tau_{\rm rec} < $ 10 years from long-period LMXBs ($P_{\rm orb} \simgt  12$~hours). 
\citet{Lin2019} attributed this distinct outbursting behavior to the critical orbital periods ($P_{\rm orb} = $ 12.4 and 12.7 hours for NS-LMXBs and BH-LMXBs, respectively), above which they can harbor {\color{black} an evolved donor}, yielding higher accretion rates, and thus shorter $\tau_{\rm rec}$.  
However, this relation between $P_{\rm orb}$ and $\tau_{\rm rec}$ is more complicated for NS-LMXBs by some selection effects, while it is more robust for BH-LMXBs, as we describe below. 

\paragraph{NS-LMXBs and recurrent transients} 
As \citet{Lin2019} pointed out, fainter X-ray outbursts may have gone undetected by all-sky monitors, and thus can bias the X-ray outburst recurrence data. This selection effect is more significant for NS-LMXBs, since the lower $L_{\rm Edd}$ yields smaller peak X-ray luminosity for a given $P_{\rm orb}$, as apparent in the \citet{Wu2010} formula. As a result, the relation between $P_{\rm orb}$ and $\tau_{\rm rec}$ is not  reliable for NS-LMXBs with $P_{\rm orb} \simlt 20$ hrs (i.e. most of the NS-LMXB sources sampled by \citet{Lin2019}) since some of their X-ray outbursts are likely too faint to be detected by all sky monitoring. This implies that a larger number of (faint) X-ray outbursts from NS-LMXBs with $P_{\rm orb} \simlt 20$ hrs could have gone undetected by all-sky monitors \citep{Lin2019} -- this is particularly true for NS-LMXBs at large distances such as those in the GC. 

However, the outburst history of X-ray transients in the GC has been robustly determined by frequent X-ray monitoring over  the last two decades -- all six NS-LMXBs within $\sim50$~pc of the GC  have short recurrence time ($\simlt5$~years) on average. A good example is \axj, with an 8.4 hr orbital period \citep{Ponti2017a},  which has shown multiple X-ray outbursts in the last decade. 
{\color{black} Therefore, we conclude that the quiescent XRBs in the GC are most likely {\it not} NS-LMXBs since no X-ray outbursts have been detected from any of them over the last two decades. } 
We emphasize that this argument is robust as it is based solely on the unbiased historical data of the X-ray transients in the GC. 

\paragraph{BH-LMXBs and single X-ray outbursters} 
In contrast to the NS-LMXB case,  \citet{Lin2019} argued that the relation between $P_{\rm orb}$ and $\tau_{\rm rec}$ is robust for local BH-LMXBs since their relatively brighter X-ray outbursts are not subject to the same selection effects, except for ultra-compact BH-LMXBs with $P_{\rm orb} \simlt 0.9$ hrs (which are likely VFXTs as mentioned below). All BH-LMXBs with $P_{\rm orb} < 12$~hours listed by \citet{Lin2019} are SXOTs, aside from the BH-LMXB XTE~J1118$+$480 with $P_{\rm orb} = 4.08$ hours. 
{\color{black} Note that XTE~J1118$+$480 was marginally considered a VFXT, since its peak X-ray luminosity  barely exceeded $L_{\rm X} = 10^{36}$~erg\,s$^{-1}$ (i.e. the conventional threshold between VFXTs and "regular" X-ray transients) and that its orbital period is just above the period gap \citep{Maccarone2013}. XTE~J1118$+$480 is an outlier in several respects and likely belongs to a unique class of BH-VFXTs.}   
Thus, we conclude that the 12 non-thermal X-ray sources, with no X-ray outbursts detected in the past two decades, are likely BH-LMXBs with $P_{\rm orb} \simlt 12$~hrs. 
Additionally, as three of the four SXOTs in the GC have the peak $L_X \simgt 10^{37}$~erg\,s$^{-1}$ corresponding to $L_X/L_{\rm Edd} \simgt 1$\% (for a 10$M_\odot$ BH), their orbital periods should be longer than $P_{\rm orb} \sim 4$ hrs, following the \citet{Wu2010} formula given above. Indeed, one of the SXOTs (CXOGC J174540.0$-$290031, whose outburst occurred in 2005) has a 7.9 hr orbital period \citep{Porquet2005}.

\begin{deluxetable*}{lccccc}[b]
\tablecaption{Classification of LMXBs in the GC}
\tablecolumns{6}
\tablehead{ \colhead{Source} &\colhead{\# of sources}   & \colhead{Location}  & 
\colhead{Recurrence time}  & \colhead{Compact object type} &  \colhead{Orbital period range}}
\startdata
Non-thermal sources & 12  & $r \simlt 1$~pc & $\simgt 13$~yrs & BH\tablenotemark{a} & $\sim 4-12$ hrs\tablenotemark{a}    \\
SXOTS (single X-ray outbursters)  & 4  & $r \simlt 2$~pc & $\simgt 13$~yrs & BH\tablenotemark{a} &  7.9 hrs\tablenotemark{c} or $\sim 4-12$ hrs\tablenotemark{a} \\
NS-LMXBs & 6 & $r \simgt 3$~pc & $\simlt 5$~yrs & NS & 8.35 hrs\tablenotemark{d} otherwise unknown \\ 
Recurrent VFXTs & 5 & $r < 10$~pc & $\simlt 10$~yrs & Unknown & Unknown \\ 
VFXTs with single outbursts & 4 & $r \simgt 10$~pc  & $\simgt 10$~yrs\tablenotemark{b} & Unknown & Unknown \\ 
\enddata
\tablenotetext{a}{These are our estimated parameter ranges or source types based on the analysis in this paper, not the measured values.}
\tablenotetext{b}{It is possible that only single outbursts have been detected because faint X-ray outbursts may have been missed by \swift/XRT due to the large off-axis distances. The actual recurrence time could be shorter like those VFXTs at $r < 10$~pc. }
\tablenotetext{c}{CXOGC~J174540.0$-$290031 \citep{Porquet2005}.}  
\tablenotetext{d}{\axj\  \citep{Maeda1996}.}  
\label{tab:src_types}
\end{deluxetable*}

\subsubsection{Quiescent X-ray emission of the dozen   non-thermal X-ray sources and the single X-ray outburst transients} \label{sec:quiescent}

Our follow-up analysis of the 2012 ACIS-S/HETG data, presented in \S\ref{sec:chandra}, firmly establishes that the 12 X-ray sources emit non-thermal X-rays and that at least 9 of them are variable, indicating that a majority are qXRBs. The combined ACIS-I and ACIS-S spectra  for the individual sources, replacing the hardness ratios (H18), yield a more accurate measurement of the photon indices. 
The mean of their photon indices, listed in Table~\ref{tab:13sources}, is  {$\Gamma = 2.2\pm0.2$}, which also matches with the best-fit photon index determined from the stacked \chandra\ ACIS-I spectra of the non-thermal X-ray sources in H18.  This is consistent with the mean photon index ($\Gamma \approx 2$) of quiescent BH-LMXB sources in the Galactic Plane   \citep{Plotkin2013, Reynolds2014}.

\begin{figure}
\begin{center} 
\includegraphics[width=8.5cm]{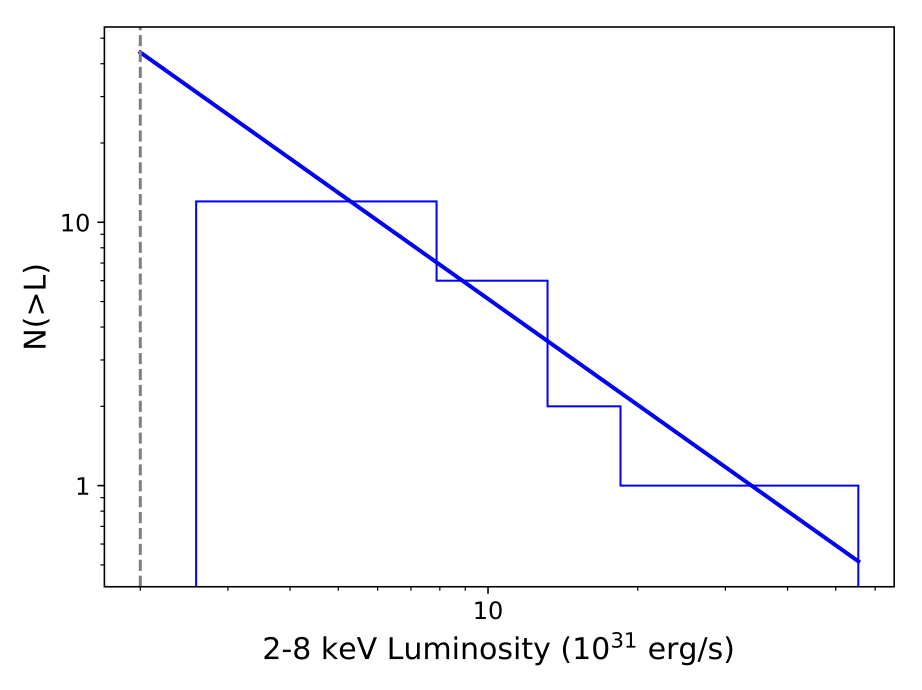}
\caption{2--8~keV luminosity function of the 12 non-thermal sources. We adopted the mean $L_{\rm X}$ value for each source listed in Table~\ref{tab:13sources}. The luminosity function is fit by a power-law model  ($N(> L_{\rm X}) \propto L_{\rm X}^{-\alpha}$) with $\alpha = 1.3 \pm 0.1$. The upper limit of quiescent $L_{\rm X}$ for the two \swift\ transients (which are likely outbursting BH-LMXBs) detected in 2016 is shown as a dashed vertical line at $L_{\rm X} = 2\times10^{31}$~erg\,s$^{-1}$. 
\label{fig:logN-logL}}
\end{center} 
\end{figure}

\citet{Padilla2014} and originally \citet{Garcia2001} found that BH-LMXBs are generally fainter than NS-LMXBs in quiescence, despite some dependence on the orbital period, distance uncertainty, and some contribution from soft thermal emission in the case of NS-LMXBs. According to \citet{Padilla2014}, in the range of  $P_{\rm orb} \approx 4-12$ hours, BH-LMXBs and NS-LMXBs have $L_X \sim 1\times10^{30}-4\times10^{31}$~erg\,s$^{-1}$ and $L_X \sim 10^{32}-10^{34}$~erg\,s$^{-1}$, respectively. 
After converting the 2--8 keV luminosity values listed in Table \ref{tab:13sources} to the 0.5--10~keV band, the $L_{\rm X}$ range of the 12 non-thermal sources ($5.6\times10^{30}$ to  $1.6\times10^{33}$~erg\,s$^{-1}$, with the mean value at $2.8\times10^{32}$~erg\,s$^{-1}$) falls somewhere between those of the BH-LMXBs and NS-LMXBs compiled by \citet{Padilla2014}. 
Figure~\ref{fig:logN-logL} shows the 2--8~keV luminosity function of the 12 non-thermal sources fit by a power-law model. Given some source variability, we adopted the mean values between the ACIS-I, ACIS-S/HETG and ACIS-S observations using the luminosity data listed in Table~\ref{tab:13sources}. We determined the best-fit slope $\alpha = 1.3\pm0.1$ where $N(>L_X) \propto  L_X^{-\alpha}$.  {Fitting to only the 9 sources with variability over 90\% CL yields $\alpha =  1.2 \pm 0.1$.}

However, there are several reasons why the 12 sources may well represent a brighter tail of the intrinsic luminosity distribution. Firstly, there are other, fainter X-ray sources  {in the central pc} which have not yet been  spectroscopically identified due to their lower X-ray counts {; we have only analyzed the brighter among the X-ray sources}. Typically, we need $\simgt 100$ \chandra\ ACIS counts (2--8 keV) for hardness ratio analysis and $\simgt 300$ counts for spectral fitting in order to distinguish non-thermal X-ray sources from CVs (H18). Secondly, there are local, quiescent BH-LMXBs at luminosities below the faintest of the 12 non-thermal sources, as shown in  \citet{Padilla2014}{, indicating that BH-LMXBs of similarly low X-ray luminosities may lie undetected at the GC}.  Thirdly, some X-ray transients have not been detected in quiescence and their $L_{\rm X}$ upper limits are much lower than those of the 12 sources. For example, we determined that $L_X \simlt  2\times10^{31}$~erg\,s$^{-1}$ in quiescence for the two \swift\ transients detected in 2016 \citep{Mori2019}. 
Given that the true luminosity distribution  {of this source population} should be skewed to a lower $L_X$ range for the above reasons  {(and our assumption that the 12 non-thermal sources are representative of the same population)}, we argue that the underlying  source population of the 12 non-thermal sources and the four SXOTs is more consistent with  {the luminosity distribution of} BH-LMXBs. In summary, the distribution of the photon indices and quiescent $L_{\rm X}$ of the 12 sources supports the BH-LMXB scenario, in addition to the X-ray outburst history discussed in \S\ref{sec:outburst_history}. All of the source types and their classifications discussed above are listed in Table~\ref{tab:src_types}.

\subsection{Nature of VFXTs in the Galactic Center}
\label{sec:vfxt_nature}

It is generally accepted that VFXTs have smaller accretion rates that produce fainter X-ray outbursts in the range of $L_X = 10^{34}$--$10^{36}$~erg\,s$^{-1}$. There are two plausible candidates for the VFXT source type. NS-LMXBs with strong magnetic fields can inhibit accretion flow via propeller effects, thus reducing the mass accretion significantly \citep{Heinke2015,Eijnden2018}. Alternatively, ultra-compact XRBs with very short orbital periods ($P_{\rm orb} \simlt 1$ hour) can accommodate only hydrogen poor companions due to the small binary separation, and thus the smaller Roche lobe size leads to much weaker accretion flow \citep{Hameury2016}.
By applying the DIM to hydrogen poor companions, \citet{Hameury2016} showed theoretically that ultra-compact XRBs can produce recurrent, faint, short-duration ($\sim$ weeks) X-ray outbursts over a few years. These outburst features in the DIM match with those of the VFXTs in general. 
As \citet{Maccarone2013} pointed out, some VFXTs may be LMXBs in the so-called period gap, with 2--3 hour orbital periods, where their convective companion stars do not fill their Roche lobes thus reducing the mass accretion rates. 

Table~\ref{tab:src_types} summarizes the various source types of XRBs in the GC, including VFXTs. 
Only in the central 10~pc region, where the \swift-XRT sensitivity is high enough for detecting VFXTs down to $L_X \sim 10^{34}$~erg\,s$^{-1}$, all 5 VFXTs have $\simlt 10$~yr recurrence time on average.   {The 4 VFXTs located outside of the central 10~pc have each been detected in outburst once.  It is uncertain if that is due to an intrinsically longer recurrence time for that population, or simply an observational bias effect resulting from decreased sensitivity.} 

Alternatively, transient intermediate polars (IPs) could account for a fainter population of VFXTs ($L_{\rm X} \sim 10^{34}$~erg\,s$^{-1}$) such as XMMU J175035.2$-$293557, which was discovered in the Galactic Bulge \citep{Hofmann2018}. However, the 9 VFXTs in the GC are not likely IPs since their X-ray outburst  {$L_{\rm X}$} increased by a factor of over 100, compared to this highly variable ($\sim20\times$) IP. 
The nature of the VFXTs in the GC, whether they are ultra-compact XRBs (with a BH or NS) or NS-LMXBs with truncated accretion disks, remains an open question, at least  {unless/}until their orbital periods are measured or type I X-ray bursts are detected during future recurrent X-ray outbursts. {\color{black} An ongoing \swift-XRT survey of the Galactic Bulge with follow-up optical/IR observations suggested that XRBs with evolved donors may also contribute to the VFXT population \citep{Shaw2020}. } 


\section{Spatial distribution of the black hole binaries}  \label{sec:bh_distribution} 

In this section we report our investigation of the spatial distribution of BHBs in the GC. We amalgamated the 12 quiescent, non-thermal X-ray sources detected by our \chandra\ data analysis (\S\ref{sec:chandra}); two \swift\ transients from 2016 \citep{Mori2019}; the transient CXOGC~J174540.0$-$290031 \citep{Muno2005b}; and the transient 1A~1742$-$289 \citep{Davies1976}, for a total of 16 BH-LMXB  candidates (BHCs hereafter). In this section, we assume that all 12  non-thermal X-ray sources are BHCs. { Nevertheless, we also performed the same analysis and present the results by excluding the three sources with no significant ($>$ 90\% CL)  variability detection.}  All errors quoted in this section correspond to 1-$\sigma$ uncertainty.

\begin{figure*}[t]
\begin{center}
\includegraphics[width=6.5cm]{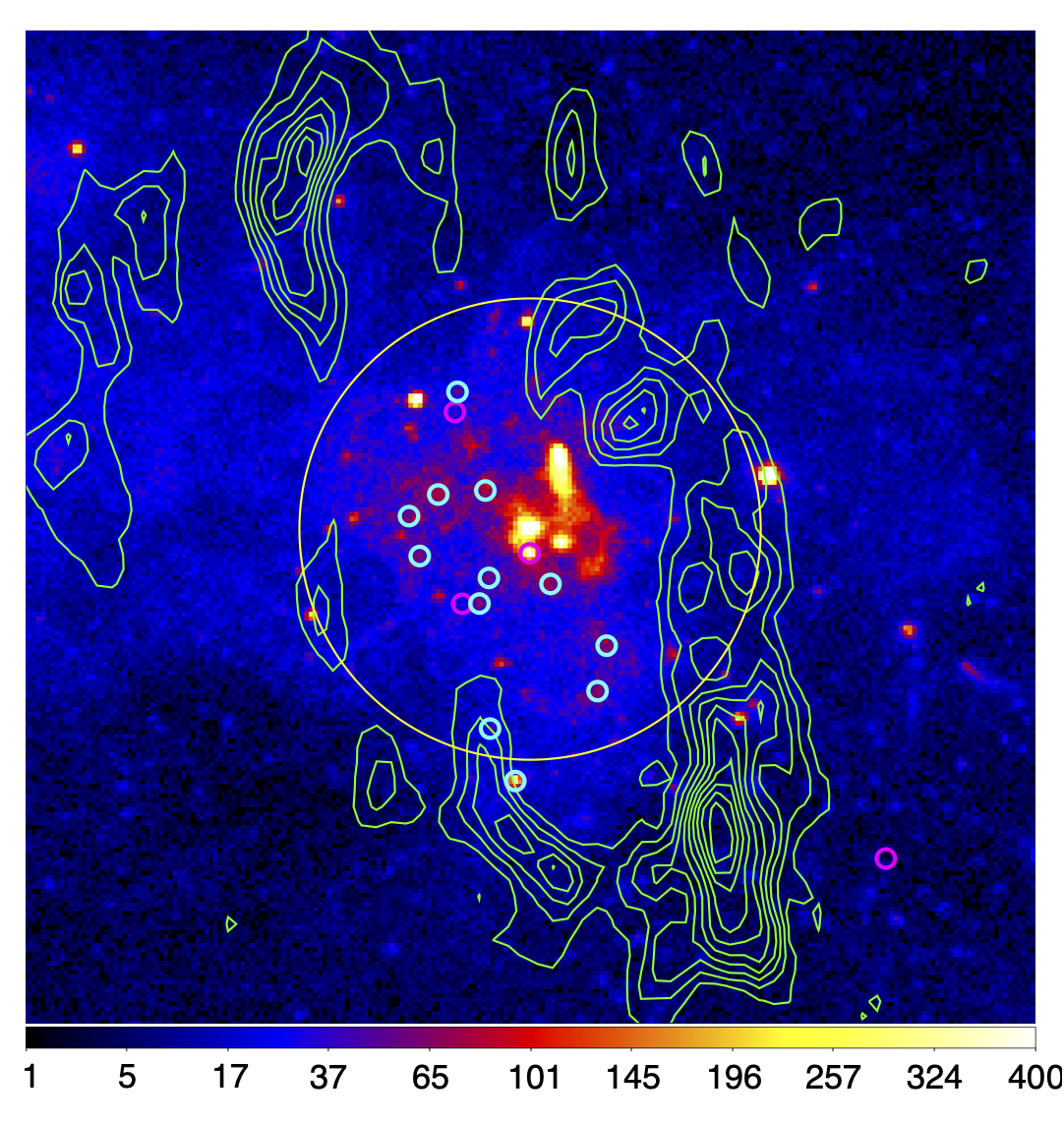}
\hspace{0.3cm}
\includegraphics[width=7.5cm]{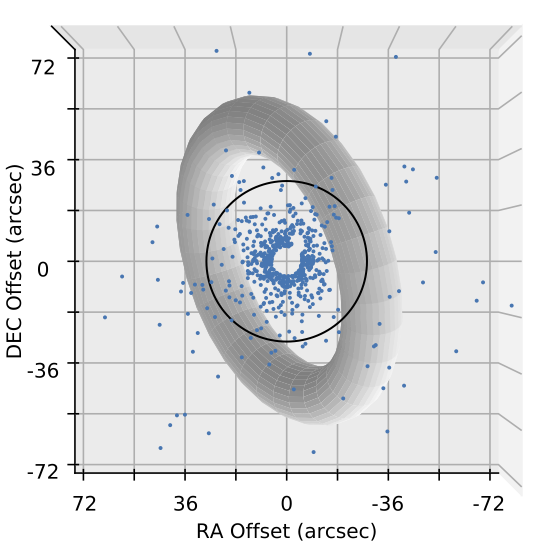}

\caption{\footnotesize Left: 2--8~keV \chandra\ ACIS-I image centered on Sgr A* of the 12 quiescent BH-LMXBs (cyan) and four BH transients (i.e., SXOTs, magenta) with CND contours (green). Right: A plot of 1,000 simulated sources (blue) overlaid with the 3D CND torus is shown to illustrate their spatial distribution and obscuration. Note that the regions at $r<0.2$~pc from Sgr A* and overlapping the bright diffuse X-ray sources (as shown in the left panel) are devoid of simulated sources. 
Also, some of the simulated sources inside/behind the CND are removed as they are not observable due to the significant X-ray  absorption.  The radius of the circle around Sgr A* in each figure is 1 pc, corresponding to $r=25$\asec.
\label{fig:chandra_images}
}
\end{center}
\end{figure*}

\subsection{Spatial fitting of the black hole binary distribution}

The 16 BHCs are projected on the 2-8~keV \chandra\ ACIS-I image in the left panel of Figure~\ref{fig:chandra_images}. The mean centroid of the 16 BHCs is offset toward the south-east direction from Sgr A* by $\Delta$RA = 0.8\asec\ and $\Delta$DEC = -7.6\asec. The overall spatial distribution seems to be elongated along the south-east direction. 
First, we employed a linear regression to the positions of the 16 BHCs to characterize their 2D spatial distribution. Our linear regression fit yields a $-8\pm3\asec$ DEC offset from Sgr A* and a $38^{+7}_{-9}$ [deg] position angle (defined relative to the north pole in the equatorial coordinate system).  

Then, we fit a 2-D Gaussian profile
\begin{equation}
   N(x,y) = e^{a(x-x_0)^2-b(x-x_0)(y-y_0)-c(y-y_0)^2} 
\end{equation}
where 
\[a={\cos^2(\theta)}/{2\sigma^2_x} + {\sin^2(\theta)}/{2\sigma^2_y}\]
\[b = {\sin(2\theta)}/{2\sigma^2_x} - {\sin(2\theta)}/{2\sigma^2_y}\]
\[c = {\sin^2(\theta)}/{2\sigma^2_x} + {\cos^2(\theta)}/{2\sigma^2_y}\]
to the 16 source  positions, using the maximum likelihood estimation (MLE) method. $\theta$ is the position angle. $(x_0, y_0)$ are the coordinates of the centroid of the Gaussian ellipse as offset from Sgr A*. $\sigma_x$ and $\sigma_y$ are the widths along the major and minor axis, respectively. The best-fit centroid ($\Delta$RA = $1\pm3$\asec\ and $\Delta$DEC = $-8\pm3$\asec) is nearly consistent with the position of Sgr A*. The best-fit position angle ($\theta$) is $41^{+11}_{-10}$ [deg]. These values are consistent with the linear regression results. The widths along the semi-major and semi-minor axis are $17\asec{}^{+5}_{-4}$ and $7\asec{}^{+2}_{-1}$, respectively, yielding the eccentricity ($e \equiv \sqrt{1-\sigma^2_y/\sigma^2_x}$) of $e = 0.91^{+0.05}_{-0.20}$.   {Fitting to only the 9 sources with variability above 90\% CL yields all parameter values within error of these reported values.} We fit the 16 sources to a de-projected 3D radial power-law profile, $n(r) = kr^{-\gamma}$. The best-fit values are $k=0.13^{+0.13}_{-0.06}$ arcsec$^{-2}$ and $\gamma = 2.3^{+0.3}_{-0.2}$. 

The NSC is well-aligned with the Galactic Plane and therefore has a position angle of $31.4^\circ\pm0.1^\circ$ \citep{Schoedel2014a}. On the other hand, the position angle of the clockwise young stellar disk in the central parsec is 99$^{\circ}\pm3^{\circ}$ \citep{Bartko2009, Lu2009}. Hence, the spatial distribution of the 16 BHCs is aligned with the NSC but not with the young stellar disk. In the next sections, we will investigate whether the disk-like distribution is induced by some observational biases.

\subsection{Observational systematics due to the circumnuclear disk and bright diffuse X-ray emission}

In the central few parsecs region, the circumnuclear disk (CND) and bright diffuse X-ray emission from the pulsar wind nebula (PWN) G359.95$-$0.04, Sgr A* and IRS 13 may inhibit observation of the intrinsic population distribution of BH-LMXBs. The hydrogen column density measurements of the CND in the radio and X-ray bands vary significantly from $N_{\rm H}\sim 10^{23}$ to $10^{25}$ cm$^{-2}$ \citep{Christopher2005, Mossoux2018}. The lower end of the CND column density ($N_{\rm H}\sim10^{23}$ cm$^{-2}$) is comparable to the typical $N_{\rm H} \sim$(1--2)$\times10^{23}$~cm$^{-2}$ measured for the X-ray transients in the central parsec.  On the other hand, if $N_{\rm H} \simgt 10^{25}$ cm$^{-2}$, our XSPEC simulations show that an X-ray source with similar fluxes and spectral shapes as the observed quiescent BH-LMXBs will be completely absorbed in the \chandra\ energy band. 

In order to account for these observation biases, we adopted the most conservative assumption that any X-ray sources inside/behind the CND or overlapping with the bright diffuse emission are undetectable by \chandra. Based on the geometry presented in \citet{Zhao2016}, we modelled the CND as a 3-dimensional torus with an inclination angle of 61$^{\circ}$ and a “roll” angle of 31$^{\circ}$ to reflect the observed 19$^\circ$ position angle of the CND's semi-major axis (see the right panel of Figure~\ref{fig:chandra_images} for the CND geometry). 

\subsection{Monte Carlo simulation}

We employed Monte Carlo (MC) simulations to test the hypothesis that the observed population of 16 BHCs represents an intrinsically spherical population centered at Sgr A*. Each simulation run generated a number of point sources using a spherically symmetric distribution with a 3D radial power-law profile of $n(r)\sim r^{-\gamma}$ with $\gamma = 2.3$ extending from $r_{\rm min} = 0.2$ pc to $r_{\rm max}$ = 2.1 pc. A  $\gamma$ value was updated from the 16 BHCs, which is different from that of H18 ($\gamma = 2.4$).  $r_{\rm min}= 0.2$~pc was chosen as X-ray point sources cannot be detected within 0.2 pc of Sgr A* due to its bright diffuse emission. $r_{\rm max}$ = 2.1 pc corresponds to the most distant BHC. Each simulated source was assigned 3D coordinates with respect to Sgr A*. 
We discarded sources that are located inside the CND or behind its line-of-sight. We also filtered out sources that overlap with Sgr A*, IRS13, PWN G359.95$-$0.04, and other diffuse emission features apparent in the \chandra\ image. In each MC realization, we generated fake sources so that we ended up with a number of ``surviving" sources ($N_S$ hereafter) similar to the observed number (i.e. 16 BHCs  within $r\simlt 2$~pc).

After taking into account the observational systematics described above, we fit a 2-D gaussian function, described in equation (1), to the ``surviving" simulated sources  projected on the 2D sky plane. The best-fit position angle ($\theta$) and eccentricity ($e$) from each MC realization were determined using the same MLE method as applied to the 16 BHCs. This simulation and fitting process was iterated 10,000 times to construct an empirical probability distribution of the position angle and eccentricity.

As a result of the MC simulation, we observed that $\sim 37\%$ of the simulated sources inside $r=2$~pc are obscured either by the CND or the bright diffuse emission. Most of the simulated sources, generated from a spherical distribution, exhibit distinct deviations from the observed orientation of the 16 BHCs. The distribution of position angles of the simulated X-ray sources is nearly uniform, as predicted from an assumed spherical distribution with some obscuration effects. 
The number of sample MC runs that fall within 1-$\sigma$ of the best-fit position angle of the 16 BHCs ($\theta = 41^{+11}_{-10}$ [$^\circ$]) is 1287 out of 10,000 realizations ($\sim13$\%). Therefore, we conclude that the observed BHCs are not consistent with a spherical distribution at  87\% confidence level. On the other hand, we found that the distribution of eccentricity is highly dependent on $N_S$. When $N_S$  is large ($\simgt 100$), we found that the eccentricity distribution was peaked at $e = 0$ and decreases toward $e=1$, in line with the predicted outcome from spherical source distribution. In contrast, a smaller number of $N_S$ ($\simlt 20$) led to a distribution skewed toward a higher $e$ value (e.g., $e = 1$ in the most extreme case when $N_S = 2$). Therefore, due to the limited number of observed samples (i.e. 16 BHCs) biasing the eccentricity distribution, we did not investigate the elongation of the source distribution using $e$. 

Alternatively, we conducted a KS test to investigate whether the BHC distribution is statistically consistent with a spherically symmetric population. The KS test is non-parametric thus it is complementary to the 2D gaussian fit described above. First, we converted the 16 BHB positions to polar coordinates ($r$, $\theta$) around Sgr A* and computed the cumulative distribution of their 16 position angles ($\theta$). Second, we constructed a cumulative distribution function (CDF) from the observable point sources in the above MC simulation; this serves as a model CDF. We found that 10,000 MC-simulated point sources deviate slightly from the uniform distribution over $\theta$ due to the obscuration and absorption effects in the GC. The model CDF was fit to an analytic function so that it could be implemented as an input to the KS test. As a result, a KS test on the actual CDF of 16 BHCs and the model CDF yields $K = 0.28$ corresponding to a p-value of 0.11. Both the parametric and non-parametric tests suggest that the hypothesis of a spherical distribution is rejected at $\sim 90$\% CL.

Our further simulations show that we need at least 20 more BHCs in the central 1 pc to firmly establish whether BHCs are  distributed around Sgr A* like a disk at $>3\sigma$ level using their PA measurements. Future detection of BH transients by \swift\ and identifying more quiescent BH-LMXBs from the archived \chandra\ ACIS-S data analysis will enhance the number of BH-LMXBs in the central parsec region and allow us to determine the spatial distribution of the BHB cusp more accurately. The {\it Lynx} X-ray observatory, given its sub-arcsecond angular resolution and significantly larger effective area than \chandra, will be best suited for detecting and identifying a large number of fainter qXRBs in the GC.

\section{Discussion} \label{sec:discussion}

Both the spatial and luminosity distributions of the 16 BHCs have several important implications for the recent XRB formation models in the GC as described below. 

\subsection{Spatial distribution of BH binaries}

As shown in Figure~\ref{fig:gc_transients} and \ref{fig:chandra_images}, it is apparent that most of the 16 BHCs are concentrated in the central 1~pc, while the six NS-LMXBs are spread out over $\sim50$~pc from the GC. 
Note that \citet{Bortolas2017} predicted that most NS should escape to outside the NSC due to their natal kicks \citep{Bortolas2017}. BH and WD are well populated within the NSC but the BH distribution is more cusp-like near the GC due to mass segregation, with a 3D power-law slope of $\gamma = 1.72\pm0.04$ at $r < 1.4$~pc whereas $\gamma = 1.00\pm0.03$ for WD \citep{Panamarev2019}. These pictures are consistent with the high concentration of 16 BHCs inside $r\simlt1$~pc. The updated power-law slope of the observed BHC  distribution ($\gamma = 2.3^{+0.3}_{-0.2}$) is close to  the model prediction ($\gamma = 1.72\pm0.04$). Furthermore, \citet{Tagawa2020} demonstrated that gas-capture binaries (i.e. binaries formed by dissipating the kinetic energy in a dense gas) can reproduce the concentration of the 16 BHCs, by applying their AGN disk model to binary formation in the Galactic Center. 

Although it is still only marginally significant, the disk-like distribution of the 16 BHCs has implications for recent theoretical predictions. \citet{Szolgyen2018} predicted that the isolated BH distribution should be elongated due to vector resonant relaxation. In this model, there is no specific orientation of the BH disk but it is probably determined by the latest episode of star formation in the GC. Our results indicate that the 16 BHCs are aligned with the NSC. Since the XRB formation in the GC follows the capture of companion stars in the NSC, the BH-LMXB distribution may track the NSC closely \citep{Generozov2018,Panamarev2019}. Using the resonance friction mechanism developed by \citet{Rauch1996},  \citet{Gruzinov2020} also found that the distribution of stellar-mass BHs should  be elongated and aligned with a rotating star cluster around the supermassive BH. 

\subsection{Number of BH-LMXBs in the central parsec region}

By extending the updated X-ray luminosity function of the 12 BHCs to a lower $L_{\rm X}$ value and using the best-fit slope $\alpha = 1.3\pm0.1$ (where $N(>L_X) \propto  L_X^{-\alpha}$), we can estimate the total number of qBH-LMXBs in the central parsec region.  
Note that the  quiescent $L_X$ upper limits of the 2016 \swift\ transients (i.e., two of the four SXOTs) indicate that fainter BH-LMXBs exist in the central parsec below $L_X \approx   2\times10^{31}$~erg\,s$^{-1}$ \citep{Mori2019}. It is apparent that the number estimate is most sensitive to the uncertainty in $\alpha$, and especially $L_{\rm min}$. By extending the best-fit X-ray luminosity function to $L_{\rm min} = 2\times10^{31}$~erg\,s$^{-1}$, we set the minimum number of BH-LMXBs to $\sim 45$ for $\alpha = 1.3$ or $\sim$ 40--50 by allowing the range of $\alpha = 1.2$--1.4.  
The number of BH-LMXBs increases if we adopt a previously reported luminosity distribution of local BH-LMXBs. Among  the quiescent BH-LMXBs and their X-ray luminosities listed in \citet{Padilla2014}, the faintest BH-LMXB (SWIFT~J1357.2$-$0933, $L_{\rm X} = 8\times10^{29}$~erg\,s$^{-1}$)  may belong to a distinct group of BH-VFXTs and thus possibly follow a different luminosity function. Also, its distance uncertainty may lead to a higher quiescent X-ray luminosity of $L_{\rm X} = 1.3\times10^{31}$~erg\,s$^{-1}$ \citep{Padilla2014}. Excluding the potential outlier SWIFT~J1357.2$-$0933, the other BH-LMXBs in the Galactic plane  have quiescent $L_X$ values above  $\sim3\times10^{30}$~erg\,s$^{-1}$. By adopting $L_{\rm min} = 3\times10^{30}$~erg\,s$^{-1}$, we deduce that $\sim 500$--630 BH-LMXBs should populate the central parsec. However, if the three sources with no significant variability detection are rMSPs, the number of BH-LMXBs will be reduced to  {$\sim$240--300}. These numbers, which we consider as the most accurate estimate, are lower than those in H18 ($\sim$ 600--1,000 and $\sim$ 450--750 BH-LMXBs, respectively) since H18 adopted $L_{\rm min} = 8\times10^{29}$~erg\,s$^{-1}$ from  \citet{Padilla2014} as well as a different $\alpha$ value of 1.8.  

\citet{Panamarev2019} predicted that $\sim$6,000 isolated BHs should reside inside 1.4~pc after 5~Gyrs and that the fraction of binaries should reach $\sim$2.5--5\%, indicating $\sim$150--300 BH binaries. \citet{Generozov2018} predicted that $\sim$60--200 BH binaries should be formed by tidal capture. These numerical results are roughly consistent with the number of BH-LMXBs estimated from the measured luminosity function of the 12 non-thermal sources. Also, as discussed in \S\ref{sec:vfxt_nature}, some of the VFXTs may represent a fainter population of BH-LMXBs (with orbital periods shorter than a few hours) which would increase the number of BH binaries in
our estimates. It is thus imperative to identify the nature of VFXTs in the GC and detect fainter, quiescent BH-LMXBs in the solar neighborhood.  

\subsection{Population of close BH binaries in the Galactic Center}

Our systematic studies of the X-ray outburst history suggest that the 16 BHCs  should have orbital periods in the range of $\sim4$--12 hrs. The N-body simulation of \citet{Panamarev2019} predicts that the binary separation (i.e. semi-major axis of binaries) is skewed toward $a\sim0.01$ [AU] after 5 Gyrs since the close binary systems have a higher chance of surviving against gravitational collisions with  other stars in the NSC or 3-body interaction with the SMBH at Sgr A*. The binary separation between two stars with mass $m_1$ and $m_2$ is given by  $a=2.3\times10^{-3}(m_1+m_2)^{1/3}P^{2/3}_{\rm orb}$ [AU] where $P_{\rm obs}$ is an orbital period [hrs]. In \S\ref{sec:gc_srcs_nature}, we argued that the 16 BHCs should have $P_{\rm orb} \sim $4--12 hrs. For a 10 $M_\odot$ BH binary  with $P_{\rm orb} = 4$--12 hrs, $a \sim $0.01--0.03 [AU]. This range is consistent with the semi-major axis distance distribution of binaries in the  inner NSC region as derived from the N-body simulation \citep{Panamarev2019}. In addition, \citet{Generozov2018} also argued that XRBs should have orbital periods shorter than $\sim10$~hours if they are formed via tidal capture around Sgr A*. Both our study of X-ray sources in the GC and the recent theoretical models for XRB formation around Sgr A* agree that BH binaries in the central parsec should be tightly bound with $P_{\rm orb} \simlt 12$ hours. If future X-ray observations reveal that some of the VFXTs in the GC are ultra-compact XRBs with $P_{\rm orb}$ shorter than a few hours, it will support this picture even further.

\section{Summary} \label{sec:summary} 

\begin{itemize} 

\item Our subsequent analysis of the 3~Msec \chandra\ ACIS-S/HETG and 1.6~Msec ACIS-S data from 2012--2018 confirmed 11 of the dozen non-thermal X-ray sources from H18 (which was based on 1.4~Msec ACIS-I data) and detected an additional non-thermal source in the central parsec as a result of much improved statistics. We detected significant long-term variability from 9 of the 12 non-thermal sources, indicating that a majority of them are quiescent XRBs rather than rMSPs. { It is still possible that one and three of the dozen non-thermal sources may be an SS Cyg-like CV or rMSPs, respectively. }  

\item Based on the most complete X-ray outburst data collected by frequent X-ray monitoring of the GC, we found that the 12 quiescent XRB candidates (with no outburst detected) as well as the four X-ray transients with single outbursts are distinct from the six NS-LMXBs and five VFXTs (within $r\simlt 10$~pc) which display recurrent X-ray outbursts. 

\item The 16 non-recurring XRBs - 12 quiescent, non-thermal sources and 4 SXOTs - are likely BH-LMXBs with an orbital period range of $P_{\rm orb} \sim 4-12$~hours, based on their outburst history and quiescent X-ray properties as inferred from the recent studies of X-ray outbursts by \citet{Lin2019} and \citet{Wu2010}. 

\item The spatial distribution of the 16 BH-LMXB candidates is consistent with a disk-like distribution aligned with the NSC. A  spherical distribution is marginally rejected at a the $\sim90$\% level. 

\item An updated X-ray luminosity function yields the number of BH-LMXBs in the central parsec $\sim500-630$, { or $\sim240-300$ if the three sources with no significant variability detection are rMSPs}. These estimates are roughly consistent with the predictions by the theoretical model and N-body simulation employed by  \citet{Generozov2018} and \citet{Panamarev2019}, respectively. 

\item While only six outbursting NS-LMXBs have been identified through type I X-ray bursts within $r \simlt 50$~pc, a larger number of quiescent NS-LMXBs may exist in the GC as observed in globular clusters  \citep{Heinke2003}.  Those faint NS-LMXBs with mostly soft thermal X-ray emission ($kT \simlt 0.1$~keV) would remain undetected as their X-ray emission is heavily obscured by the large hydrogen column density ($N_{\rm H} \sim 10^{23}$~cm$^{-2}$) in the GC, and they do not show X-ray outbursts (not even faint ones) due to their low accretion rates. 

\item We derived a binary separation range of $a\sim0.01-0.03$~[AU] of the BH-LMXB population in the central parsec, assuming that a majority of them have $P_{\rm orb} \sim 4-12$~hours and $10M_\odot$ BH mass. These results provide direct evidence for the predictions of \citet{Generozov2018} and  \citet{Panamarev2019}. Both our results and their models suggest that XRBs in the central parsec should be tightly bound; otherwise, the gravitational interaction with the SMBH at Sgr A* and collisions with other stars in the NSC can destroy the binaries. 

\end{itemize}


\acknowledgments

We thank Brian Metzger, Nicholas Stone, Aleksey Generozov, Barry McKernan, K.E. Saavik Ford and Mordecai-Mark Mac Low for discussions on the population of X-ray binaries and gravitational wave events in galactic nuclei. {\color{black} We thank Benjamin Vermette for simulating \chandra\ ACIS spectra of non-magnetic CVs, including SS~Cyg.} GP acknowledges financial contribution from the agreement ASI-INAF n.2017-14-H.0.  

\software{HEASoft Version 6.27 \citep{heasoft}, FTools Version 6.27 \citep{heasoft}}

\end{document}